\documentclass[superscriptaddress,secnumarabic,nobibnotes,aps,prd,showkeys,noshowpacs,onecolumn,12pt]{revtex4}
\usepackage{graphics}
\usepackage{graphicx}
\usepackage{color}
\usepackage{epsf}
\usepackage{bm}
\usepackage{amsmath,amssymb,amsfonts,amsthm}
\usepackage{latexsym}
\usepackage{enumerate}
\usepackage{hyperref}

\begin{document}

\title{Revise the Dark Matter-Phantom Scalar Field Interaction}
\author{Andronikos Paliathanasis}
\email{anpaliat@phys.uoa.gr}
\affiliation{Institute of Systems Science, Durban University of Technology, Durban 4000,
South Africa}
\affiliation{Departamento de Matem\'{a}ticas, Universidad Cat\'{o}lica del Norte, Avda.
Angamos 0610, Casilla 1280 Antofagasta, Chile}
\affiliation{School of Technology, Woxsen University, Hyderabad 502345, Telangana, India}
\author{Amlan Halder}
\email{amlankanti.halder@woxsen.edu.in}
\affiliation{School of Technology, Woxsen University, Hyderabad 502345, Telangana, India}
\author{Genly Leon}
\email{genly.leon@ucn.cl}
\affiliation{Departamento de Matem\'{a}ticas, Universidad Cat\'{o}lica del Norte, Avda.
Angamos 0610, Casilla 1280 Antofagasta, Chile}
\affiliation{Institute of Systems Science, Durban University of Technology, Durban 4000,
South Africa}

\begin{abstract}
The cosmological history and evolution are examined for gravitational models
with interaction in the dark sector of the universe. In particular, we
consider the dark energy to be described by a phantom scalar field and the
dark matter $\rho _{m}$ as a pressureless ideal gas. We introduce the
interacting function $Q=\beta \left( t\right) \rho_{m}$, where the function $%
\beta \left( t\right) $ is considered to be proportional to $\dot{\phi},~%
\dot{\phi}^{2}H^{-1},~H$, or a constant parameter with dimensions of $\left[
H_{0}\right] $. For the four interacting models, we study in details the
phase space by calculating the stationary points. The latter are applied to
reconstruct the cosmological evolution. Compactified variables are essential
to understand the complete picture of the phase space and to conclude about
the cosmological viability of these interacting models. The detailed
analysis is performed for the exponential potential $V\left( \phi \right)
=V_{0}e^{\lambda \phi }$. The effects of other scalar field potential
functions on the cosmological dynamics are examined.
\end{abstract}

\maketitle

\section{Introduction}

\label{sec1}

At the present, the universe is under an accelerated phase \cite%
{rr1,Teg,Kowal,Komatsu,suzuki11} driven by dark energy \cite{jo}. The nature
of the dark energy is unknown, but observations indicate that dark energy
introduces repulsive-gravitational forces in the universe which lead to the
cosmic expansion. The dark energy model has been addressed by cosmologists
with various proposed solutions \cite{de01,de02,de03,de04,de05}.
Nevertheless, the problem of the cosmological tensions \cite{ht1} has opened
new directions for the study of the dark energy problem \cite{ht2}.

Cosmological models with interaction within the dark sector of the universe
have been introduced \cite{Amendola:1999er}, in order to address the
coincidence problem \cite{con1,con2,con3}. However, interacting models are
applied to answer problems with the cosmological tensions \cite%
{ht3,ht4,ht5,ht6,ht7}. In the interacting models, there is energy transfer
between the elements of the dark sector of the universe, which leads to the
introduction of new terms in the continuity equations so as to have new
behaviors in the cosmic evolution. Interaction components can follow from
fundamental gravitation theory \cite{ref1,ch1,ch2,sf3,ph1}, or they are
introduced phenomenologically \cite{ss4,ss5,ss7,ss9,ss11,ss12,ss16,ss17}.
Recently, in \cite{ss18}, interacting models have been discussed which lead
to compartmentalization and coexistence in the dark sector of the universe.

In this study, we consider the dark energy to be described by a phantom
scalar field minimally coupled to gravity. The main characteristic of the
phantom scalar field is that it violates the weak energy condition, allowing
it to have negative energy density and a value for the equation of state
parameter smaller than that of the cosmological constant. However, the
equation of state parameter for the phantom scalar field model can cross the
phantom divide line only once \cite{q14}. The cosmological dynamics for the
phantom field without any interaction term leads to the late-time attractor,
which can describe a super-exponential universe culminating in a Big Rip or
other kinds of sudden singularities \cite{q15}. Nevertheless, the Big Rip
singularity can be avoided in the presence of a nonzero interaction term 
\cite{var00}.

We introduce a family of interacting functions between the dark matter and
the phantom scalar field and perform a detailed analysis of the phase-space
of the field equations in order to understand the effects of the interaction
on the cosmological dynamics and to infer the cosmological viability of the
models \cite{dn1,dn2,dn3,dn4,dn5}. For the interacting function, we consider
those provided by theories derived from the variational principle as well as
models proposed phenomenologically. The plan of the paper is as follows.

In Section \ref{sec2} we discuss the interactions within the components of
the dark sector of the universe. For a spatially flat FLRW geometry we
assume that the dark energy is described by a phantom scalar field and the
dark matter by a pressureless ideal gas. The interaction is considered to be
proportional to the energy density of the dark energy, that is, $Q=\beta
\left( t\right) \rho _{m}$. The coefficient function $\beta \left( t\right)
~ $ has the same dimensions as the Hubble parameter, and we consider four
different cases (A) $\beta \left( t\right) =\beta _{0}\dot{\phi}$,~(B) $%
\beta \left( t\right) =\beta _{0}\frac{\dot{\phi}^{2}}{H}$, (C) $\beta
\left( t\right) =\beta _{0}H$ and (D)~$\beta \left( t\right) =\beta
_{0}H_{0} $, which lead to four different interacting models.

Section \ref{sec3} includes the main analysis of this study, where we make
use of the Hubble normalization approach to study the phase-space for the
four interacting models and the exponential potential. Due to the nature of
the constraints, we employ compactified variables. From the determination of
the stationary points and the study of the asymptotic solutions at those
points, we are able to reconstruct the cosmological history and the
evolution of the physical parameters for each model. From this analysis, we
draw conclusions about the cosmological viability of the above interacting
models. In Section \ref{sec4}, we extend our discussion to a scalar field
potential function beyond the exponential, where we show that new stationary
points may exist, and the physical properties of the corresponding solutions
are consistent with the exponential potential. Finally, in Section \ref{sec5}%
, we draw our conclusions.

\section{Cosmological Interactions}

\label{sec2}

On very large scales, the universe is considered to be isotropic and
homogeneous, described by the spatially flat FLRW geometry 
\begin{equation}
ds^{2}=-N^{2}\left( t\right) dt^{2}+a^{2}(t)\left(
dx^{2}+dy^{2}+dz^{2}\right) \;.  \label{metric1}
\end{equation}%
where $a\left( t\right) $ is the scale factor that defines the radius of the
universe and $N\left( t\right) $ is a lapse function.

For the comoving observer $u^{\mu }=\frac{1}{N}\delta _{t}^{\mu }$,~$u^{\mu
}u_{\mu }=-1$, we define the expansion rate $\theta =\frac{1}{3}u_{;\mu
}^{\mu }$; that is, $\theta =\frac{1}{3}H$, where $H=\frac{1}{N}\frac{\dot{a}%
}{a}$ is the Hubble function with $\dot{a}=\frac{da}{dt}$. In the following
without loss of generality, we assume that the lapse function is a constant, 
$N\left( t\right) =1$, such that $H=\frac{\dot{a}}{a}$.

The cosmological fluid inherits the symmetries of the background spacetime
and is a perfect fluid described by the energy-momentum tensor 
\begin{equation}
T_{\mu \nu }=\rho u_{\mu }u_{\nu }+ph_{\mu \nu },
\end{equation}%
where $\rho $ is the energy density of the cosmological fluid, $p$ is the
pressure component, and $h_{\mu \nu }=g_{\mu \nu }+u_{\mu }u_{\nu }$ is the
projection tensor.

Within the framework of General Relativity, the cosmological field equations
are 
\begin{equation}
G_{\mu \nu }=T_{\mu \nu }
\end{equation}%
in which $G_{\mu \nu }=R_{\mu \nu }-\frac{R}{2}g_{\mu \nu }$ is the Einstein
tensor.

For the FLRW line element (\ref{metric1}) the components of the
gravitational field equations are%
\begin{eqnarray}
3H^{2} &=&\rho ,  \label{g.05} \\
-2\dot{H}-3H^{2} &=&p.  \label{g.06}
\end{eqnarray}%
Furthermore, the equation of motion for the cosmological fluid $T_{~~;\nu
}^{\mu \nu }=0$ leads to the differential equation%
\begin{equation}
\dot{\rho}+3H\left( \rho +p\right) =0.  \label{g.07}
\end{equation}

In terms of the deceleration parameter $q=-1-\frac{\dot{H}}{H^{2}}$, the
equation of motion (\ref{g.06}) reads%
\begin{equation}
q=\frac{1}{2}\left( 1+3w_{tot}\right) ,
\end{equation}%
where $w_{tot}=\frac{p}{\rho }$ is the equation of state parameter for the
cosmological fluid.

The cosmological fluid consists of different elements, such as the baryonic
matter, radiation, and the two components of the dark sector of the
universe: dark energy and dark matter. According to the cosmological data,
the dark sector of the universe constitutes approximately 97\% of the
universe.

We focus on the analysis of the dark sector of the universe. Thus,
initially, we assume that the cosmological fluid consists only of the dark
matter $T_{\mu \nu }^{m}$ and the dark energy $T_{\mu \nu }^{DE}$
components, such that 
\begin{equation}
T_{\mu \nu }=T_{\mu \nu }^{m}+T_{\mu \nu }^{DE}.  \label{g.08}
\end{equation}%
A pressureless fluid source describes dark matter, that is, dust, such hat 
\begin{equation}
T_{\mu \nu }^{m}=\rho _{m}u_{\mu }u_{\nu }.  \label{g.09}
\end{equation}%
while a perfect fluid describes the dark energy 
\begin{equation}
T_{\mu \nu }^{DE}=\left( \rho _{d}+p_{d}\right) u_{\mu }u_{\nu }+p_{d}g_{\mu
\nu },~  \label{g.10}
\end{equation}%
with a negative equation of the state parameter $w_{d}=\frac{p_{d}}{\rho _{d}%
}$, that is, $w_{d}<0$ to describe repulsive gravitational forces and cosmic
acceleration.

The energy density and pressure components for the cosmological fluid are
expressed as $\rho =\rho _{m}+\rho _{d}$ and $p=p_{d}$. Therefore, the
equation of motion for the cosmological fluid, i.e., the continuity equation
(\ref{g.07}) reads%
\begin{equation}
\left( \dot{\rho}_{m}+3H\rho _{m}\right) +\left( \dot{\rho}_{d}+3H\left(
\rho _{d}+p_{d}\right) \right) =0.  \label{gg.11}
\end{equation}%
Equation (\ref{gg.11}) can be written in the equivalent form \cite%
{Amendola:1999er}%
\begin{eqnarray}
\dot{\rho}_{m}+3H\rho _{m} &=&Q,  \label{g.14} \\
\dot{\rho}_{d}+3H\left( \rho _{d}+p_{d}\right) &=&-Q.  \label{g.15}
\end{eqnarray}%
where function $Q\left( t\right) $ describes the energy transfer between the
two fluid components. Indeed, for the positive valued function $Q\left(
t\right) >0$, there is a mass transfer from the dark energy fluid to dark
matter; on the other hand, for a negative-valued $Q\left( t\right) <0$, the
mass transfer is from the dark matter to the dark energy component.

\subsection{Phantom scalar field}

The dark energy component is assumed to be described by a phantom scalar
field, such that the energy density and pressure components are defined as 
\begin{eqnarray}
\rho _{d} &=&-\frac{1}{2}\dot{\phi}^{2}+V\left( \phi \right) ,  \label{g.16}
\\
p_{d} &=&-\frac{1}{2}\dot{\phi}^{2}-V\left( \phi \right) .  \label{g.17}
\end{eqnarray}%
where $V\left( \phi \right) $ is the scalar field potential which defines
the scalar field mass. The phantom scalar field, by definition, can violate
the weak energy condition, because it is possible $\rho _{d}<0$. The
equation of state parameter is defined as 
\begin{equation}
w_{d}=-\frac{\frac{1}{2}\dot{\phi}^{2}+V\left( \phi \right) }{-\frac{1}{2}%
\dot{\phi}^{2}+V\left( \phi \right) },  \label{g.18}
\end{equation}%
where in contrary to the quintessence scalar field where the equation of the
state parameter is bounded, in the phantom scalar field, there is no lower
bound for parameter $w_{d}$.

The equation of motion (\ref{g.15}) for the scalar field reads%
\begin{equation}
-\dot{\phi}\left( \ddot{\phi}+3H\dot{\phi}-V_{,\phi }\right) =-Q
\label{g.19}
\end{equation}

Regarding the interacting function $Q\left( t\right) $ in the following we
consider the interaction of the form 
\begin{equation}
Q=\beta \left( t\right) \rho _{m},  \label{g.20}
\end{equation}%
where $\beta \left( t\right) $ has the dimensions of the Hubble function.
The form of the latter interaction with a scalar field has it is origin in
Weyl Integrable Spacetime, or in scalar-tensor theories under conformal
transformation. The definition of the function $\beta \left( t\right) $
indicate the nature of the interaction; that is, the interaction is local or
global, and if it is possible for the interaction term to change sign. In
the following we consider the following cases for the function $\beta \left(
t\right) $, (A) $\beta \left( t\right) =\beta _{0}\dot{\phi}$,~(B) $\beta
\left( t\right) =\beta _{0}\frac{\dot{\phi}^{2}}{H}$, (C) $\beta \left(
t\right) =\beta _{0}H$ and (D)~$\beta \left( t\right) =\beta _{0}H_{0}$.

Model (A) is inspired by the Weyl Integrable Spacetime or the Chameleon
mechanism \cite{ref1,ch1,ch2}. Indeed the function $\beta \left( t\right) $
to be proportional to the $\dot{\phi}$ is provided by the theories with
variational principle \cite{var0,var1,var2,var3}. Model (C) employs a global
interacting function, which has been used to describe the energy transfer
when the fluids have constant equation-of-state parameters \cite{ref3,ref4}.
Model (D) represents a metastable local interaction scenario \cite%
{ref5,ref6,sp1} as a local interacting model. Finally, Model (B) describes a
generalized interaction (A) case. \ 

We investigate the phase space to study the cosmological history and the
dynamical evolution of the physical parameters described by the nonlinear
gravitational field equations. Precisely, we determine the stationary points
of the gravitational field equations and analyze the physical properties of
the asymptotic solutions associated with these stationary points.
Furthermore, to reconstruct the cosmological history, we examine the
stability properties of the stationary points.

\section{Asymptotic analysis in Hubble normalization}

\label{sec3}

We introduce the dimensionless dependent variables using the Hubble
normalization approach \cite{cop1,cop2} 
\begin{equation}
x=\frac{\dot{\phi}}{\sqrt{6}H},~y=\frac{\sqrt{V\left( \phi \right) }}{\sqrt{3%
}H},~\Omega _{m}=\frac{\rho _{m}}{3H^{2}},~\lambda =\frac{V_{,\phi }}{V},
\end{equation}%
and the new independent variable $\tau =\ln a$. Parameter $\Omega _{m}$ is
the energy density for the dark matter, while the energy density for the
dark energy is defined as $\Omega _{d}=-x^{2}+y^{2}$. We assume that $H>0$.

For each of the interaction models, we rewrite the field equations (\ref%
{g.05}), (\ref{g.06}), (\ref{g.14}) and (\ref{g.19}) in the equivalent form
of an algebraic-differential system%
\begin{eqnarray}
F\left( \mathbf{x}\right) &=&0,  \label{g.21} \\
\frac{d\mathbf{x}}{d\tau }-\mathbf{G}\left( \mathbf{x}\right) &=&0,
\label{g.22}
\end{eqnarray}%
where $\mathbf{x=}\left( x,y,\lambda ,\Omega _{m}\right) ^{T}$. \ 

Equation (\ref{g.21}) is the algebraic equation, which is independent of the
interaction, that is 
\begin{equation}
1+x^{2}-y^{2}-\Omega _{m}=0.  \label{g.23}
\end{equation}%
This constraint equation is used to reduce the dimension of the dynamical
system (\ref{g.22}). For the reduced system, we determine the points $%
\mathbf{x}_{0}$, where $\mathbf{G}\left( \mathbf{x}_{0}\right) =0$.

Each stationary point describes an asymptotic solution where the
deceleration parameter is defined as%
\begin{equation}
q\left( \mathbf{x}_{0}\right) =\frac{1}{2}\left( 1-3\left(
x_{0}^{2}+y_{0}^{2}\right) \right) \text{.}  \label{g.24}
\end{equation}%
Finally, we solve the eigenvalue problem $\left\vert \frac{\partial \mathbf{G%
}}{\partial \mathbf{x}}-eI\right\vert =0$, for the linearized system around
the stationary points, such that to determine the stability properties for
the points.

At this point, it is important to remark that the variables $x^{2},~y^{2}$
and $\Omega _{m}$ are not constrained, and they can take values in all the
range of real values. By definition, we shall consider $y\geq 0$. Thus, in
order the phase-space analysis to be completed, compactified variables
should be considered.

Recently, in \cite{sp1} the asymptotic dynamics investigated for some of
these interactions, nevertheless, some constraints for the dynamical
variables have been introduced, which does not allow for the violation for
the weak energy condition, while the analysis has not been performed with
the use of compactified variables, consequently, not all possible pathogens
of the theories have been examined.

Additionally, the existence of the singular surfaces for some of these
interactions needs to be discussed, similarly in \cite{m11}. The existence
of the singular surfaces debate the viability of these models.

\subsection{Interaction A}

For the interacting model A, the field equations in terms of the
dimensionless variables are%
\begin{eqnarray}
\frac{dx}{d\tau } &=&\frac{1}{2}\left( \sqrt{6}ax^{2}-3x^{3}-3x^{2}\left(
1+y^{2}\right) +\sqrt{6}\left( \beta _{0}+\left( \lambda -\beta _{0}\right)
y^{2}\right) \right) ,  \label{g.25} \\
\frac{dy}{d\tau } &=&\frac{1}{2}y\left( \sqrt{6}\lambda x+3\left(
1-x^{2}-y^{2}\right) \right) ,  \label{g.26} \\
\frac{d\lambda }{d\tau } &=&\sqrt{6}\lambda ^{2}x\left( \Gamma \left(
\lambda \right) -1\right) ,~\Gamma \left( \lambda \right) =\frac{V_{,\phi
\phi }V}{V_{,\phi }^{2}}\text{\thinspace }.  \label{g.27}
\end{eqnarray}

Consider the exponential potential $V\left( \phi \right) =V_{0}e^{\lambda
\phi }$, such that $\Gamma \left( \lambda \right) =1$, and the dimension of
the dynamical system (\ref{g.25}), (\ref{g.26}), (\ref{g.27}) is reduced by
one.

\subsubsection{Stationary points at the finite regime}

The stationary points $A=\left( x\left( A\right) ,y\left( A\right) \right) $
for the two-dimensional dynamical system (\ref{g.25}), (\ref{g.26}) are%
\begin{eqnarray*}
A_{1} &=&\left( \sqrt{\frac{2}{3}}\beta _{0},0\right) ,~ \\
A_{2} &=&\left( \frac{\sqrt{\frac{3}{2}}}{\beta _{0}-\lambda },\frac{\sqrt{%
\beta _{0}\left( \beta _{0}-\lambda \right) -\frac{3}{2}}}{\left\vert \beta
_{0}-\lambda \right\vert }\right) , \\
A_{3} &=&\left( \frac{\lambda }{\sqrt{6}},\sqrt{1+\frac{\lambda ^{2}}{6}}%
\right) ,~ \\
A_{4}^{\pm } &=&\left( \pm i,0\right) .
\end{eqnarray*}%
Points $A_{4}^{\pm }$ are not real, consequently they are not physically
accepted. For the three real points, the energy densities $\Omega
_{m},~\Omega _{d}$ and the deceleration parameter $q$ are calculated as%
\begin{eqnarray*}
A_{1} &:&\Omega _{m}=1+\frac{2}{3}\beta _{0}^{2},~\Omega _{\phi }=-\frac{2}{3%
}\beta _{0}^{2},~q=\frac{1}{2}-\beta _{0}^{2}, \\
A_{2} &:&\Omega _{m}=\frac{3-\beta _{0}\lambda +\lambda ^{2}}{\left( \beta
_{0}-\lambda \right) ^{2}},~\Omega _{\phi }=\frac{\beta _{0}^{2}-\beta
_{0}\lambda -3}{\left( \beta _{0}-\lambda \right) ^{2}},~q=\frac{1}{2}-\frac{%
3}{2}\frac{\beta _{0}}{\left( \beta _{0}-\lambda \right) }, \\
A_{3} &:&\Omega _{m}=0,~\Omega _{\phi }=1,~q=-1-\frac{\lambda ^{2}}{2}.
\end{eqnarray*}

Moreover, the eigenvalues of the linearized system around the real
stationary points are%
\begin{eqnarray*}
A_{1} &:&-\frac{1}{2}\left( 3+2\beta _{0}^{2}\right) ,~\frac{1}{2}\left(
3-2\beta _{0}^{2}+2\beta _{0}\lambda \right) , \\
A_{2} &:&\frac{\left( 3\beta _{0}\left( 3\lambda -2\beta _{0}\right)
-3\lambda ^{2}\pm \sqrt{\Delta \left( A_{2}^{\pm }\right) }\right) }{4\left(
\beta _{0}-\lambda \right) ^{2}}, \\
A_{3} &:&-\frac{1}{2}\left( 6+\lambda ^{2}\right) ,~-3-\lambda ^{2}+\beta
_{0}\lambda .
\end{eqnarray*}%
where 
\begin{equation*}
\Delta \left( A_{2}^{\pm }\right) =\left( \beta _{0}-\lambda \right)
^{2}\left( 4\beta _{0}^{2}\left( 15+8\lambda ^{2}\right) -72-16\beta
_{0}^{3}\lambda -21\lambda ^{2}-4\beta _{0}\lambda \left( 9+4\lambda
^{2}\right) \right) .
\end{equation*}

Stationary point $A_{1}$ describes a universe where the kinetic part of the
scalar field interacts with the dark matter component. The solution
describes an accelerated universe for $\beta _{0}^{2}\geq \frac{1}{2}$,
while the de Sitter universe is recovered for $\beta _{0}^{2}=\frac{1}{2}$.
From the eigenvalues, we conclude that the point is an attractor for $%
\left\{ \beta _{0}>0,\lambda <\frac{2\beta _{0}^{2}-3}{2\beta _{0}}\right\} $
or $\left\{ \beta _{0}<0,\lambda >\frac{2\beta _{0}^{2}-3}{2\beta _{0}}%
\right\} $.

The stationary point $A_{2}$ is real and physically accepted when $\beta
_{0}\left( \beta _{0}-\lambda \right) -\frac{3}{2}>0~$and $\beta _{0}\neq
\lambda $. The asymptotic solutions describe universes where the two fluids
of the dark sector contribute to the cosmic evolution, and there exists a
nonzero interaction term. The asymptotic solution describes an accelerated
universe when $\left\{ \beta _{0}<-\frac{1}{\sqrt{2}},\frac{2\beta _{0}^{2}-3%
}{2\beta _{0}}<\lambda <-2\beta _{0}\right\} $ or $\left\{ \beta _{0}>\frac{1%
}{\sqrt{2}},-2\beta _{0}<\lambda <\frac{2\beta _{0}^{2}-3}{2\beta _{0}}%
\right\} $ and $\left\{ \beta _{0}=-\frac{1}{\sqrt{2}},\lambda =\sqrt{2}%
\right\} $ or $\left\{ \beta _{0}=\frac{1}{\sqrt{2}},\lambda =-\sqrt{2}%
\right\} $. From the analysis of the eigenvalues, we conclude that the
stationary points, when they exist, are always saddle points.

Finally, point $A_{3}$ always describes an accelerated universe dominated by
the phantom scalar field, while the de Sitter universe is recovered for $%
\lambda =0$. The stationary point $A_{3}$ is an attractor for $\lambda =0$
or for $\beta _{0}\lambda <3+\lambda ^{2}$.

\begin{figure}[tbph]
\centering\includegraphics[width=1\textwidth]{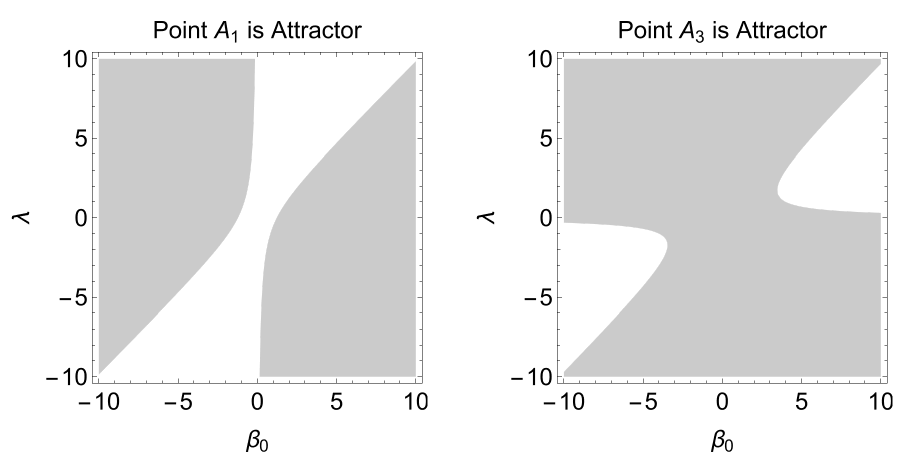}
\caption{Interaction A: Region plots in the space of variables $\left\{ 
\protect\beta _{0},\protect\lambda \right\} $ where the stationary points $%
A_{1}$ (Left Fig.) and $A_{3}~$(Right Fig.) are attractors.}
\label{ff1}
\end{figure}

In Fig. \ref{ff1} we present the regions in the space of variables $\left\{
\beta _{0},\lambda \right\} $ where the stationary points $A_{1}$ and $A_{3}$
are attractors.

\subsubsection{Compactified variables}

We have determined the stationary points within the finite regime. However,
the dynamical variables $x$ and $y$ can approach infinity, and stationary
points may also exist in this infinite regime. To address this, we introduce
compactified variables.

\begin{equation}
x=\frac{X}{\sqrt{1-X^{2}-Y^{2}}},~y=\frac{Y}{\sqrt{1-X^{2}-Y^{2}}},~dT=\sqrt{%
1-X^{2}-Y^{2}}d\tau .  \label{d.00}
\end{equation}%
where $X^{2}\leq 1$ and $Y^{2}\leq 1$ with constraint $1-X^{2}-Y^{2}\geq 0$.

Hence, for the exponential potential, the two-dimensional dynamical system (%
\ref{g.25}), (\ref{g.26}) reads%
\begin{align}
\frac{dX}{dT} &=\frac{1}{2}\left( \sqrt{6}X^{2}\left( 2\left( \beta
_{0}-\lambda \right) Y^{2}-\beta _{0}\right) +\sqrt{6}\left( \beta
_{0}+\left( \lambda -2\beta _{0}\right) Y^{2}\right) -3X\left(
1+2Y^{2}\right) \sqrt{1-X^{2}-Y^{2}}\right) ,  \label{d.01} \\
\frac{dY}{dT} &=\frac{1}{2}Y\left( 1-2Y^{2}\right) \left( 3\sqrt{%
1-X^{2}-Y^{2}}-\sqrt{6}\left( \beta _{0}-\lambda \right) X\right) .
\label{d.02}
\end{align}

The stationary points at the infinity are on the surface $1-X^{2}-Y^{2}=0$,
they are 
\begin{eqnarray*}
A_{1}^{\left( \infty \right) \pm } &=&\left( \pm 1,0\right) , \\
A_{2}^{\left( \infty \right) \pm } &=&\left( \pm \frac{1}{\sqrt{2}},\frac{1}{%
\sqrt{2}}\right) .
\end{eqnarray*}

The stationary points describe asymptotic solutions where only the phantom
field contributes the cosmological fluid, i.e., $\Omega _{m}\left(
A_{1}^{\left( \infty \right) \pm }\right) =0~$and$~\Omega _{m}\left(
A_{2}^{\left( \infty \right) \pm }\right) =0$. The stationary points
describe Big Rip singularities, that is, $q\left( A_{1}^{\left( \infty
\right) \pm }\right) =-\infty $ and $q\left( A_{2}^{\left( \infty \right)
\pm }\right) =-\infty $. Moreover, from the analysis of the eigenvalues, we
infer that the points always describe unstable solutions.

\begin{figure}[tbph]
\centering\includegraphics[width=1\textwidth]{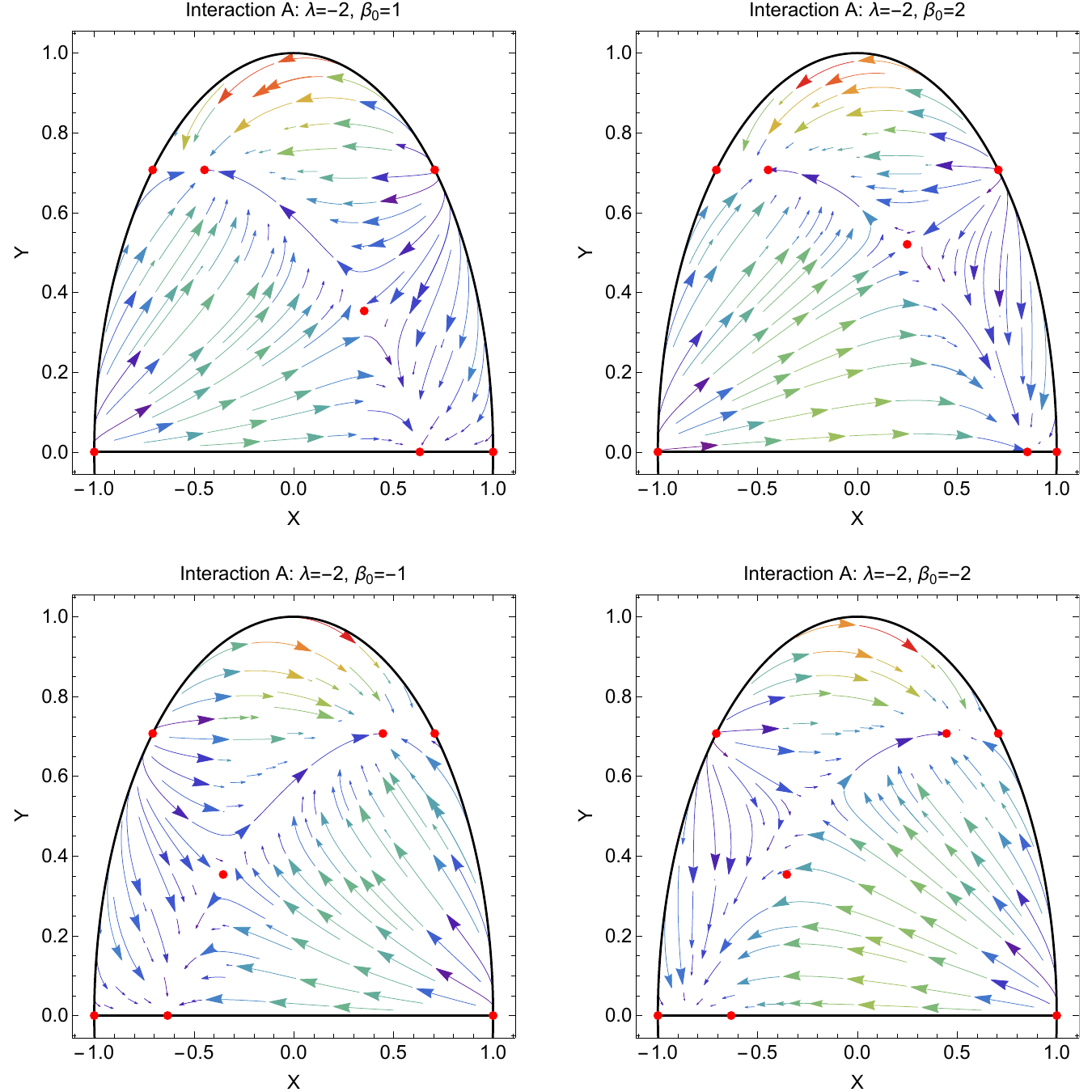}
\caption{Interaction A: Phase-space portraits for the two-dimensional
dynamical system (\protect\ref{d.01}), (\protect\ref{d.02}) for different
values of the free parameters $\left\{ \protect\beta _{0},\protect\lambda %
\right\} $. The stationary points are marked with red. We observe that
attractors exist only in the finite regime. }
\label{ff2}
\end{figure}

\begin{figure}[tbph]
\centering\includegraphics[width=1\textwidth]{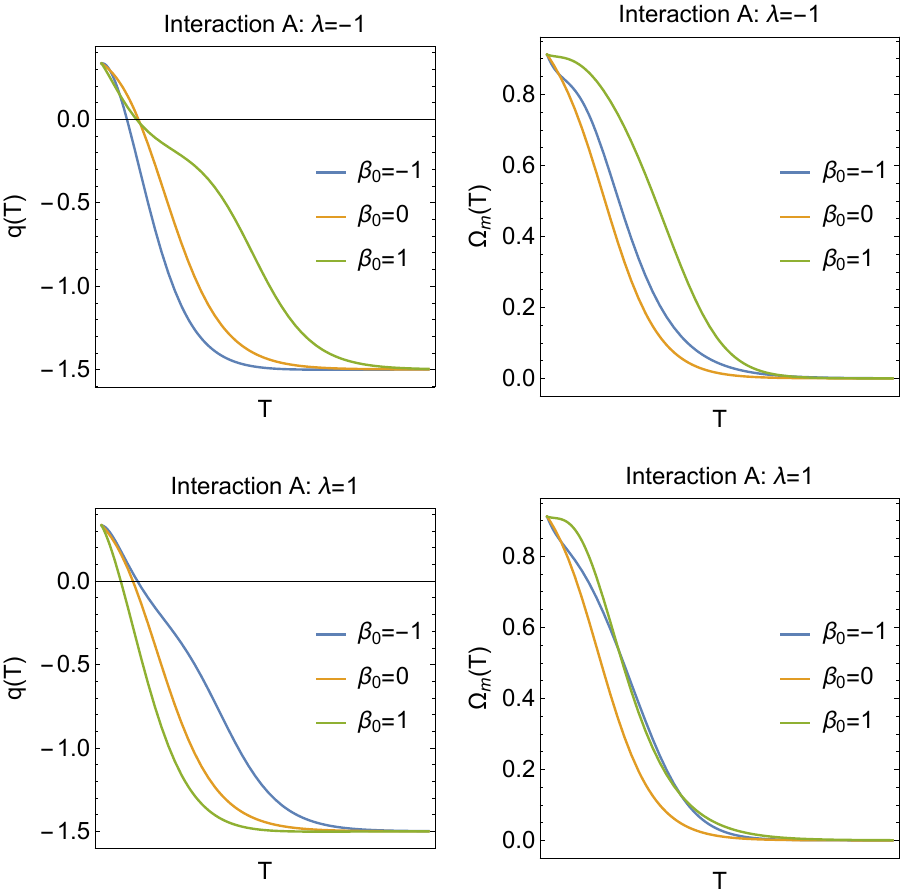}
\caption{Interaction A: Evolution for the deceleration parameter$~q$ (Left
figures) and the energy density for the dark matter $\Omega _{m}$ (Right
figures) as they are given by the numerical solution of the two-dimensional
dynamical system (\protect\ref{d.01}), (\protect\ref{d.02}) for different
values of the free parameters $\left\{ \protect\beta _{0},\protect\lambda %
\right\} $. }
\label{ff3}
\end{figure}

Phase-space portraits for the dynamical system (\ref{d.01}), (\ref{d.02})
are presented in Fig. \ref{ff2}. Moreover, in Fig. \ref{ff3}, we present
numerical evolution for the deceleration parameter $q$ and the energy
density $\Omega _{m}$.

\subsection{Interaction B}

For the interaction $Q=\beta _{0}\frac{\dot{\phi}^{2}}{H}\rho _{m}$, the
field equation in the dimensionless variables read%
\begin{eqnarray}
\frac{dx}{d\tau } &=&\frac{1}{2}\left( 3\left( 2\beta _{0}-1\right) x\left(
1+x^{2}\right) +\sqrt{6}\lambda y^{2}-3\left( 2\beta _{0}+1\right)
xy^{2}\right) ,  \label{d.03} \\
\frac{dy}{d\tau } &=&\frac{1}{2}y\left( \sqrt{6}\lambda x+3\left(
1-x^{2}-y^{2}\right) \right) ,  \label{d.04} \\
\frac{d\lambda }{d\tau } &=&\sqrt{6}\lambda ^{2}x\left( \Gamma \left(
\lambda \right) -1\right) ,~\Gamma \left( \lambda \right) =\frac{V_{,\phi
\phi }V}{V_{,\phi }^{2}}\text{\thinspace }.  \label{d.05}
\end{eqnarray}%
We continue with the analysis of the space space for the latter dynamical
system for the exponential scalar field potential where $\lambda $ is always
a constant parameter.

\subsubsection{Stationary points at the finite regime}

The stationary points $B=\left( x\left( B\right) ,y\left( B\right) \right) $
for the two-dimensional dynamical system (\ref{d.03}), (\ref{d.04}) are%
\begin{eqnarray*}
B_{1} &=&\left( 0,0\right) , \\
B_{2} &=&\left( \frac{\lambda }{\sqrt{6}},\sqrt{1+\frac{\lambda ^{2}}{6}}%
\right) , \\
B_{3} &=&\left( \frac{\lambda +\sqrt{12\beta _{0}+\lambda ^{2}}}{2\sqrt{6}%
\beta _{0}},\frac{\sqrt{\left( 2\beta _{0}-1\right) \left( 6\beta
_{0}+\lambda \left( \lambda +\sqrt{12\beta _{0}+\lambda ^{2}}\right) \right) 
}}{2\sqrt{3}\left\vert \beta _{0}\right\vert }\right) , \\
B_{4} &=&\left( \frac{\lambda -\sqrt{12\beta _{0}+\lambda ^{2}}}{2\sqrt{6}%
\beta _{0}},\frac{\sqrt{\left( 2\beta _{0}-1\right) \left( 6\beta
_{0}+\lambda \left( \lambda -\sqrt{12\beta _{0}+\lambda ^{2}}\right) \right) 
}}{2\sqrt{3}\left\vert \beta _{0}\right\vert }\right) , \\
B_{4}^{\pm } &=&\left( \pm i,0\right) .
\end{eqnarray*}%
Points $B_{4}^{\pm }$ are not real. Thus, they are not physically accepted.
For the rest of the stationary points, we calculate the physical parameters $%
\Omega _{m},~q$ as follows%
\begin{eqnarray*}
B_{1} &:&\Omega _{m}=1,~q=\frac{1}{2}, \\
B_{2} &:&\Omega _{m}=0,~q=-1-\frac{\lambda ^{2}}{2}, \\
B_{3} &:&\Omega _{m}=\frac{6\beta _{0}+\lambda \left( \lambda -\sqrt{12\beta
_{0}+\lambda ^{2}}\right) \left( 1-\beta _{0}\right) }{6\beta _{0}^{2}}%
,~q=-1-\frac{\lambda }{4\beta _{0}}\left( \lambda +\sqrt{12\beta
_{0}+\lambda ^{2}}\right) , \\
B_{4} &:&\Omega _{m}=\frac{6\beta _{0}+\lambda \left( \lambda +\sqrt{12\beta
_{0}+\lambda ^{2}}\right) \left( 1-\beta _{0}\right) }{6\beta _{0}^{2}}%
,~q=-1-\frac{\lambda }{4\beta _{0}}\left( \lambda -\sqrt{12\beta
_{0}+\lambda ^{2}}\right) .
\end{eqnarray*}

Moreover, the eigenvalues are calculated%
\begin{eqnarray*}
B_{1} &:&\frac{3}{2},\frac{3}{2}\left( 2\beta _{0}-\frac{1}{2}\right) , \\
B_{2} &:&-\frac{1}{2}\left( 6+\lambda ^{2}\right) ,~-3-\lambda ^{2}+\beta
_{0}\lambda , \\
B_{3} &:&\frac{-\beta _{0}\left( 2\beta _{0}-1\right) \lambda \left( \lambda
+\sqrt{12\beta _{0}^{2}+\lambda ^{2}}\right) \pm \sqrt{2\beta _{0}\left(
2\beta _{0}-1\right) \Delta \left( B_{3}\right) }}{8\beta _{0}^{2}}, \\
B_{4} &:&\frac{-\beta _{0}\left( 2\beta _{0}-1\right) \lambda \left( \lambda
-\sqrt{12\beta _{0}^{2}+\lambda ^{2}}\right) \pm \sqrt{2\beta _{0}\left(
2\beta _{0}-1\right) \Delta \left( B_{4}\right) }}{8\beta _{0}^{2}},
\end{eqnarray*}%
with%
\begin{eqnarray*}
\Delta \left( B_{3}\right) &=&288\beta _{0}^{2}+6\beta _{0}\lambda
^{2}\left( 20+\beta _{0}\left( 2\beta _{0}-17\right) \right) +\lambda
^{4}\left( 8+\beta _{0}\left( 2\beta _{0}-9\right) \right) \\
&&+\lambda \sqrt{12\beta _{0}+\lambda ^{2}}\left( 24\beta _{0}\left(
3-2\beta _{0}\right) +\lambda ^{2}\left( 8-\beta _{0}\left( 9-2\beta
_{0}\right) \right) \right) ,
\end{eqnarray*}%
\begin{eqnarray*}
\Delta \left( B_{4}\right) &=&288\beta _{0}^{2}+6\beta _{0}\lambda
^{2}\left( 20+\beta _{0}\left( 2\beta _{0}-17\right) \right) +\lambda
^{4}\left( 8+\beta _{0}\left( 2\beta _{0}-9\right) \right) \\
&&-\lambda \sqrt{12\beta _{0}+\lambda ^{2}}\left( 24\beta _{0}\left(
3-2\beta _{0}\right) +\lambda ^{2}\left( 8-\beta _{0}\left( 9-2\beta
_{0}\right) \right) \right) .
\end{eqnarray*}

The point $B_{1}$ describes the dark-matter-dominated epoch. For $\beta _{0}>%
\frac{1}{4}$ the point is a source; otherwise, it is a saddle point.
Furthermore, point $B_{2}$ describes a universe dominated by the scalar
field, and it has the same physical and stability properties as point $A_{3}$%
. Finally, at the stationary points $B_{3}$ and $B_{4}$, the asymptotic
solution describes a nonzero interacting term between the two fluid sources.

The stationary points $B_{3}$,~$B_{4}$ exist when $\beta _{0}\geq \frac{1}{2}
$. While cosmic acceleration is recovered by the corresponding asymptotics
solutions when $\left\{ \lambda \leq -\sqrt{2},\beta _{0}>\frac{\lambda ^{2}%
}{4}\right\} $ or $\left\{ \lambda >-\sqrt{2}\right\} $ for point $B_{3}$
and $\left\{ \lambda \geq \sqrt{2},\beta _{0}>\frac{\lambda ^{2}}{4}\right\} 
$ or $\left\{ \lambda <\sqrt{2}\right\} $ for point $B_{4}$. For $\lambda =0$%
, the de Sitter spacetime is recovered.

\begin{figure}[tbph]
\centering\includegraphics[width=1\textwidth]{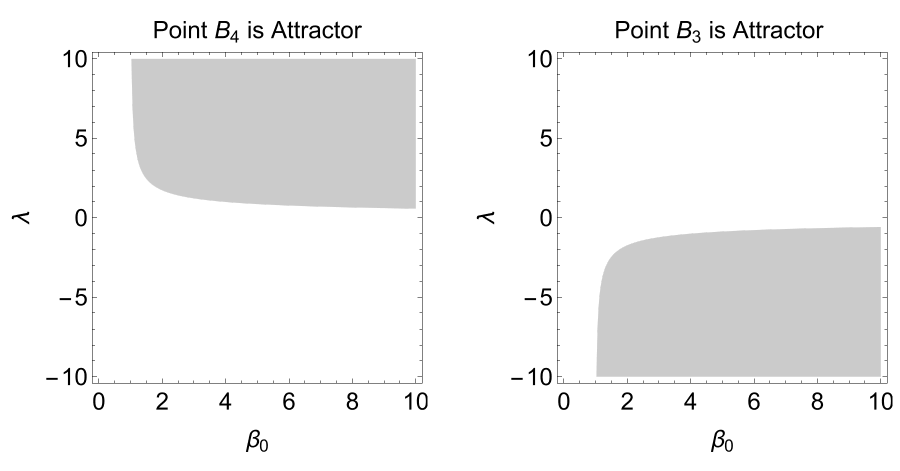}
\caption{Interaction B: Region plots in the space of variables $\left\{ 
\protect\beta _{0},\protect\lambda \right\} $ where the stationary points $%
B_{3}$ (Left Fig.) and $B_{3}~$(Right Fig.) are attractors.}
\label{ff4}
\end{figure}

Finally, in Fig. \ref{ff4}, we present the region plots in the space of the
free parameters $\left\{ \beta _{0},\lambda \right\} $ where these two
stationary points are attractors. It is important to mention that when the
points are attractors, the asymptotic solutions always describe cosmic
acceleration. 

\subsubsection{Compactified variables}

We continue with the analysis of the dynamical system (\ref{d.03}), (\ref%
{d.04}) by using the compactified variables (\ref{d.00}) to investigate the
existence of stationary points at infinity.

In terms of the compactified variables, the dynamical system (\ref{d.03}), (%
\ref{d.04}) reads%
\begin{align}
\frac{dX}{dT } &=\frac{1}{2}\left( \sqrt{6}\lambda Y^{2}\left(
1-2Y^{2}\right) -3X\left( 1+2Y^{2}\right) \sqrt{1-X^{2}-Y^{2}}+6\beta _{0}X%
\frac{\left( 1-X^{2}\right) \left( 1-2Y^{2}\right) }{\sqrt{1-X^{2}-Y^{2}}}%
\right) ,  \label{d.06} \\
\frac{dY}{dT } &=\frac{1}{2}Y\left( 1-2Y^{2}\right) \left( \sqrt{6}\lambda
X+3\sqrt{1-X^{2}-Y^{2}}+3\beta _{0}\frac{X^{2}}{\sqrt{1-X^{2}-Y^{2}}}\right)
.  \label{d.07}
\end{align}

The stationary points at infinity are 
\begin{eqnarray*}
B_{1}^{\left( \infty \right) \pm } &=&\left( \pm 1,0\right) , \\
B_{2}^{\left( \infty \right) \pm } &=&\left( \pm \frac{1}{\sqrt{2}},\frac{1}{%
\sqrt{2}}\right) .
\end{eqnarray*}%
The stationary points describe Big Rip singularities, similar to the points
at infinity for the interacting model A. Unlike in the previous case, where
stationary points at infinity are always unstable, we find that the points $%
B_{1}^{(\infty)\pm}$ are attractors for $\beta_0 > \frac{1}{2}$. In
contrast, the points $B_{2}^{(\infty)\pm}$ always correspond to unstable
solutions. We conclude that to avoid the introduction of a Big Rip
singularity as a future attractor for any set of initial conditions, the
parameter $\beta_0$ must be constrained as $\beta_0 < \frac{1}{2}$. In this
case, the stationary points $B_{3}$ and $B_{4}$ are not real.

\begin{figure}[tbph]
\centering\includegraphics[width=1\textwidth]{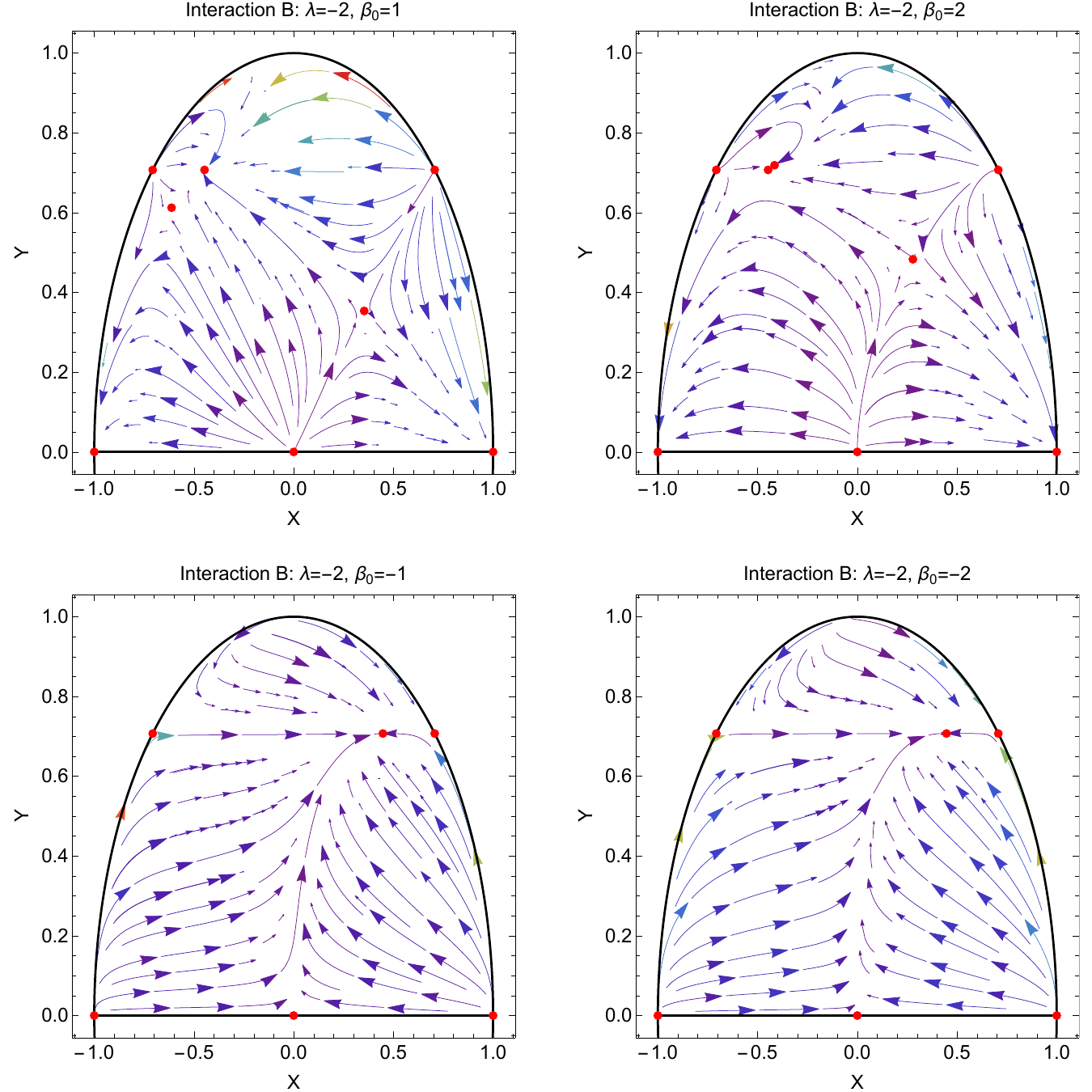}
\caption{Interaction B: Phase-space portraits for the two-dimensional
dynamical system (\protect\ref{d.06}), (\protect\ref{d.07}) for different
values of the free parameters $\left\{ \protect\beta _{0},\protect\lambda %
\right\} $. The stationary points are marked with red.}
\label{ff5}
\end{figure}

\begin{figure}[tbph]
\centering\includegraphics[width=1\textwidth]{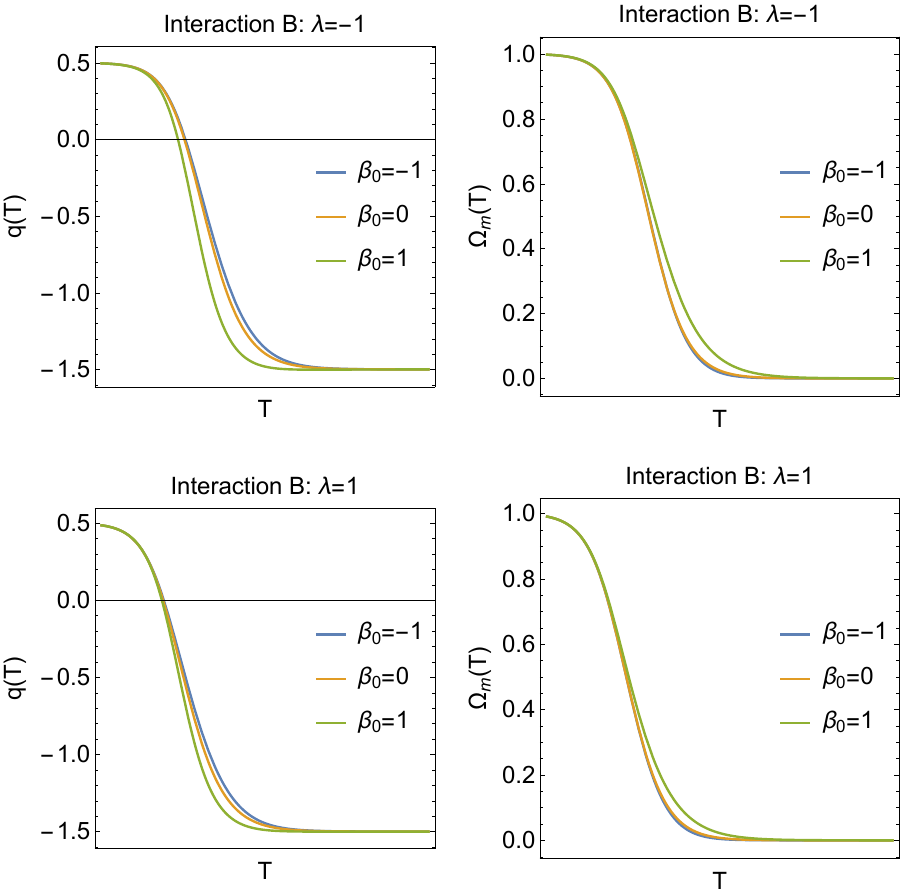}
\caption{Interaction B: Evolution for the deceleration parameter$~q$ (Left
figures) and the energy density for the dark matter $\Omega _{m}$ (Right
figures) as they are given by the numerical solution of the two-dimensional
dynamical system (\protect\ref{d.06}), (\protect\ref{d.07}) for different
values of the free parameters $\left\{ \protect\beta _{0},\protect\lambda %
\right\} $. }
\label{ff6}
\end{figure}

Phase-space portraits for the dynamical system (\ref{d.06}), (\ref{d.07})
are presented in Fig. \ref{ff5}. Furthermore, in Fig. \ref{ff6}, we show the
numerical evolution for the deceleration parameter $q$ and the energy
density $\Omega_m$. We have imposed initial conditions for the numerical
solutions where the trajectories have attractors in the finite regime.

\subsection{Interaction C}

For the third interaction model, namely $Q=\beta _{0}H\rho _{m}\,$\ the
field equations expressed in the dimensionless variables are 
\begin{eqnarray}
\frac{dx}{d\tau } &=&\frac{1}{2}\left( \sqrt{6}\lambda y^{2}+3\left(
1+x^{2}+y^{2}+\beta _{0}\frac{1+x^{2}-y^{2}}{2x}\right) \right) ,
\label{d.08} \\
\frac{dy}{d\tau } &=&\frac{1}{2}y\left( \sqrt{6}\lambda x+3\left(
1-x^{2}-y^{2}\right) \right) ,  \label{d.09} \\
\frac{d\lambda }{d\tau } &=&\sqrt{6}\lambda ^{2}x\left( \Gamma \left(
\lambda \right) -1\right) ,~\Gamma \left( \lambda \right) =\frac{V_{,\phi
\phi }V}{V_{,\phi }^{2}}\text{\thinspace }.  \label{d.10}
\end{eqnarray}

For the exponential potential where $\Gamma \left( \lambda \right) =1$ and $%
\lambda $ is always a constant, the stationary points $C=\left( x\left(
C\right) ,y\left( C\right) \right) $ for the latter dynamical system are%
\begin{eqnarray*}
C_{1}^{\pm } &=&\left( \pm \sqrt{\frac{\beta _{0}}{3}},0\right) , \\
C_{2} &=&\left( \frac{\beta _{0}-3}{\sqrt{6}\lambda },\frac{\sqrt{2\beta
_{0}\lambda ^{2}-\left( \beta _{0}-3\right) ^{2}}}{\sqrt{6}\left\vert
\lambda \right\vert }\right) , \\
C_{3} &=&\left( \frac{\lambda }{\sqrt{6}},\sqrt{1+\frac{\lambda ^{2}}{6}}%
\right) , \\
C_{4}^{\pm } &=&\left( \pm i,0\right) .
\end{eqnarray*}

For points $C_{1}^{\pm }$,~$C_{2}$ and $C_{3}$ we calculate the physical
parameters%
\begin{eqnarray*}
C_{1}^{\pm } &:&\Omega _{m}=1+\frac{\beta _{0}}{3},~q=\frac{1-\beta _{0}}{2},
\\
C_{2} &:&\Omega _{m}=\frac{\left( 3-\beta _{0}\right) \left( 3-\beta
_{0}+\lambda ^{2}\right) }{3\lambda ^{2}},~q=\frac{1-\beta _{0}}{2}, \\
C_{3} &:&\Omega _{m}=0,~q=-1-\frac{\lambda ^{2}}{2}.
\end{eqnarray*}%
Moreover, we calculate the eigenvalues%
\begin{eqnarray*}
C_{1}^{\pm } &:&-\left( \beta _{0}+3\right) ,~\frac{1}{2}\left( 3-\beta
_{0}\pm \sqrt{2\beta _{0}}\lambda \right) , \\
C_{2} &:&\frac{2\beta _{0}\lambda ^{3}-\left( \beta _{0}-3\right) \left(
\beta _{0}+1\right) \lambda \pm \sqrt{\Delta \left( C_{2}\right) }}{4\left(
\beta _{0}-3\right) \lambda ^{2}}, \\
C_{3} &:&-\frac{1}{2}\left( 6+\lambda ^{2}\right) ,~-3-\lambda ^{2}+\beta
_{0}\lambda .
\end{eqnarray*}%
where%
\begin{eqnarray*}
\Delta \left( C_{2}\right) &=&8\left( \beta _{0}-5\right) ^{2}-3\left( \beta
_{0}-7\right) \left( \beta _{0}-3\right) ^{2}\left( 5\beta _{0}-3\right)
\lambda ^{2} \\
&&~~+4\left( \beta _{0}-15\right) \left( \beta _{0}-3\right) \beta
_{0}\lambda ^{4}+4\beta _{0}^{2}\lambda ^{6}.
\end{eqnarray*}

Stationary points $C_{1}^{\pm }$ are real when$~\beta _{0}>0$; the
asymptotic solution describe cosmic acceleration when $\beta _{0}>1$. From
the corresponding eigenvalues, we conclude that the points are attractors
when $\pm \lambda <\frac{\beta _{0}-3}{\sqrt{2\beta _{0}}}$.

\begin{figure}[tbph]
\centering\includegraphics[width=1\textwidth]{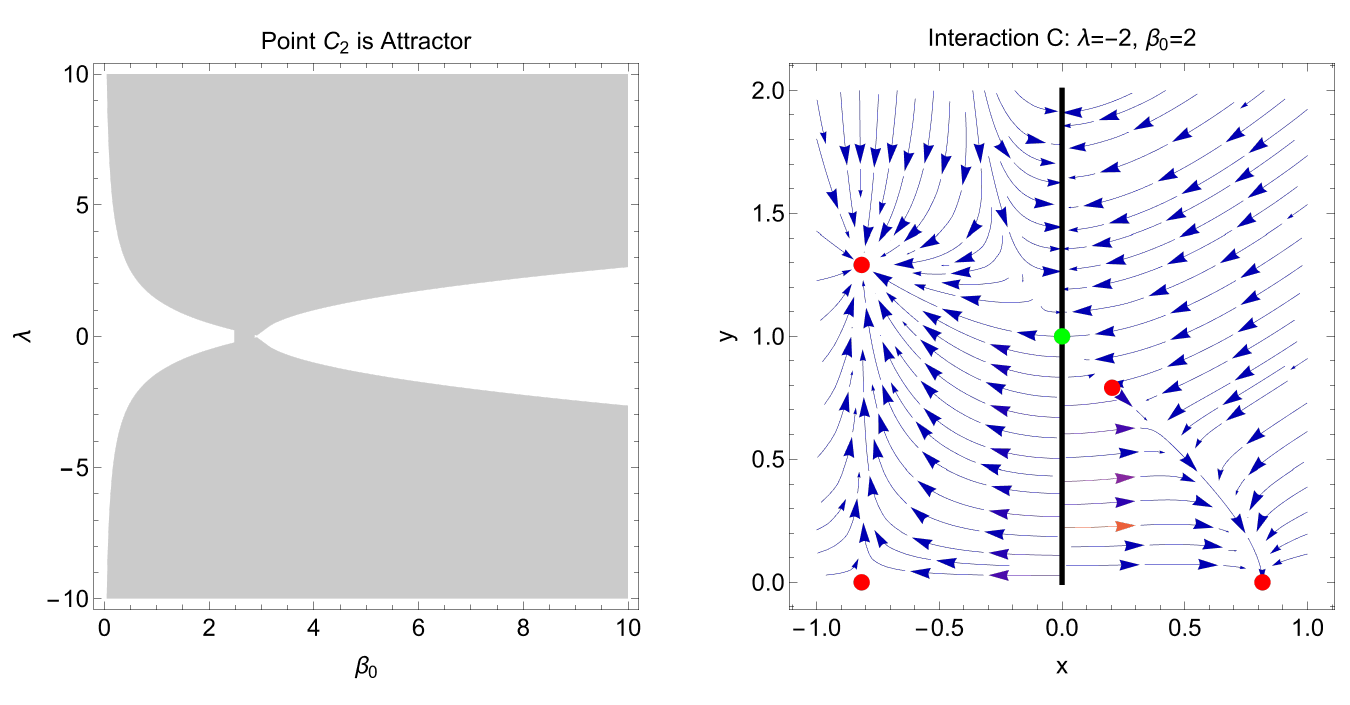}
\caption{Interaction C: Left figure: Region plot in the space $\left\{ 
\protect\beta _{0},\protect\lambda \right\} $ where point $C_{2}$ is an
attractor. Right figure: Phase-space portrait for the dynamical system (%
\protect\ref{d.08}), (\protect\ref{d.09}). Red marks the stationary points,
while green marks the transition point $C_{T}$. The black line defines the
singular points with $x=0$. }
\label{ff7}
\end{figure}

Point $C_{2}$ is real when $\beta _{0}>0$ and $\lambda ^{2}>\frac{1}{2}%
\left( \frac{9}{\beta _{0}}+\beta _{0}-6\right) $,~$\lambda \neq 0$. The
asymptotic solutions describe a universe with a nonzero interaction term;
acceleration is recovered when $\beta _{0}>1$. In Fig. \ref{ff7}, we present
the region in the space of the free variables $\left\{ \lambda,\beta
_{0}\right\} $ where $C_{2}$ is an attractor. Finally, point $C_{3}$ has the
same physical and stability properties with points $A_{3}$ and $B_{2}$.

However, the dynamical system (\ref{d.08}), (\ref{d.09}) possess a
singularity at $x=0$. The point $C_{T}=\left( 0,1\right) $ exists, which is
not a stationary point but a transition point that allows the trajectories
to move into solutions with different signs for variable $x$. In Fig. \ref%
{ff7} we present the phase-space for the dynamical (\ref{d.08}), (\ref{d.09}%
) where we show the transition of the solutions as they move near to $C_{T}$%
. The family of points $x=0$ behaves as sources for $y<1$ and attractors for 
$y>1$. This behaviour is for the positive value of parameter $\beta _{0}>0$.
Nevertheless, for $\beta _{0}<0$ the points on the line $x=0$, behaves as
sources for $y>1$ and like attractors for $y<1.$

\subsubsection{Compactified variables}

We investigate the existence of stationary points at infinity by working in
the compactified variables (\ref{d.00}). The dynamical system (\ref{d.08}), (%
\ref{d.09}) is written as follows%
\begin{align}
\frac{dX}{dT} &=\frac{1}{2}\left( \sqrt{6}\lambda Y^{2}\left(
1-2Y^{2}\right) -\left( 3X\left( 1+2Y^{2}\right) -\frac{\beta _{0}\left(
1-X^{2}\right) \left( 1-2Y^{2}\right) }{X}\right) \sqrt{1-X^{2}-Y^{2}}%
\right) ,  \label{d.11} \\
\frac{dY}{dT} &=\frac{1}{2}Y\left( 1-2Y^{2}\right) \left( \sqrt{6}\lambda
X+3\left( 1-\beta _{0}\right) \sqrt{1-X^{2}-Y^{2}}\right) .  \label{d.12}
\end{align}%
The stationary points on the surface $1-X^{2}-Y^{2}=0$, are%
\begin{eqnarray*}
C_{1}^{\left( \infty \right) \pm } &=&\left( \pm 1,0\right) , \\
C_{2}^{\left( \infty \right) \pm } &=&\left( \pm \frac{1}{\sqrt{2}},\frac{1}{%
\sqrt{2}}\right) .
\end{eqnarray*}%
which describe Big Rip singularities. The analysis of the eigenvalues leads
to the conclusion that the stationary points $C_{1}^{\left( \infty \right)
\pm }$ and $C_{2}^{\left( \infty \right) \pm }$ are saddle points.

\begin{figure}[tbph]
\centering\includegraphics[width=1\textwidth]{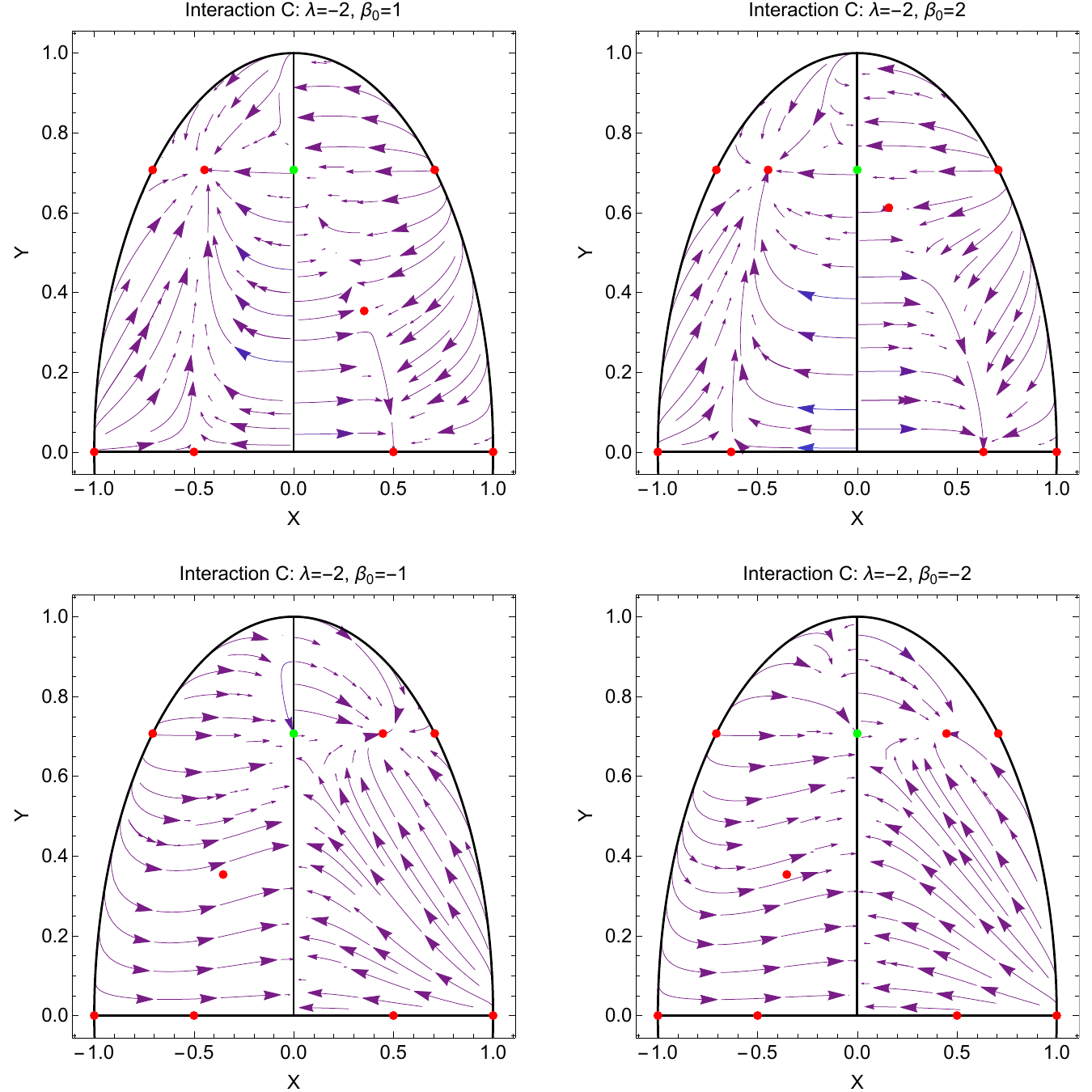}
\caption{Interaction C: Phase-space portraits for the two-dimensional
dynamical system (\protect\ref{d.11}), (\protect\ref{d.12}) for different
values of the free parameters $\left\{ \protect\beta _{0},\protect\lambda %
\right\} $. The stationary points are marked with red. With green it is
marked the transition point $C_{T}$}
\label{ff8}
\end{figure}

\begin{figure}[tbph]
\centering\includegraphics[width=1\textwidth]{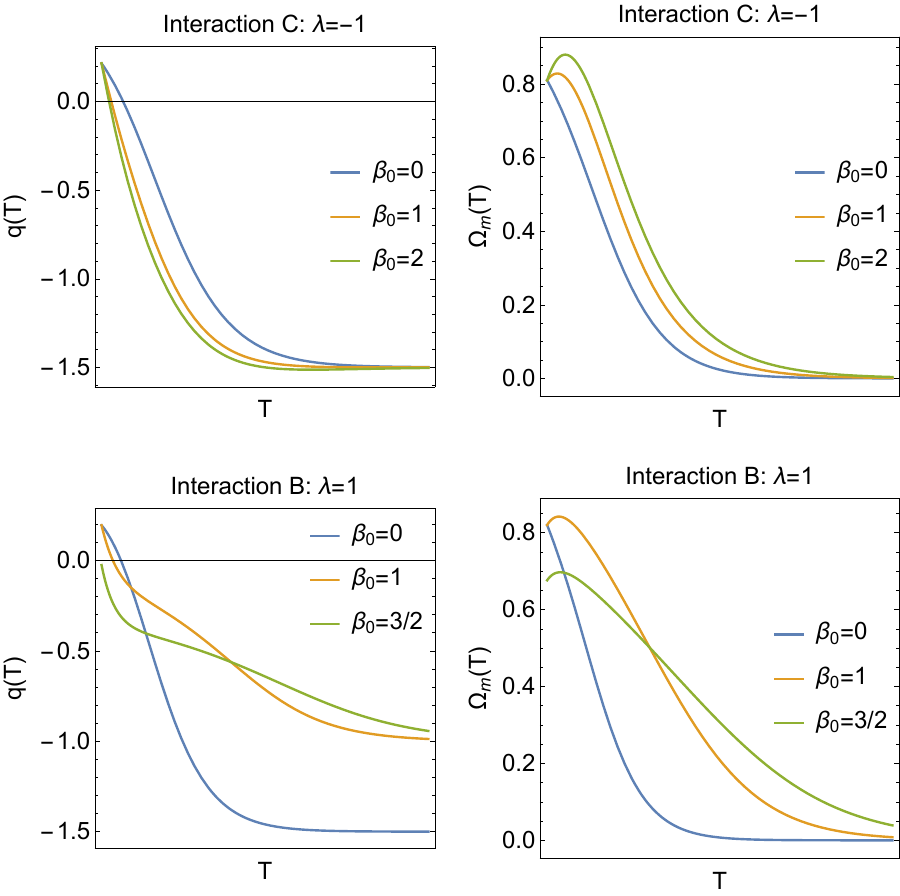}
\caption{Interaction C: Evolution for the deceleration parameter$~q$ (Left
figures) and the energy density for the dark matter $\Omega _{m}$ (Right
figures) as they are given by the numerical solution of the two-dimensional
dynamical system (\protect\ref{d.11}), (\protect\ref{d.12}) for different
values of the free parameters $\left\{ \protect\beta _{0},\protect\lambda %
\right\} $. }
\label{ff9}
\end{figure}

Phase-space portraits for the dynamical system (\ref{d.11}), (\ref{d.12})
are presented in Fig. \ref{ff8}. Moreover, in Fig. \ref{ff9}, we present
numerical evolution for the deceleration parameter $q$ and the energy
density $\Omega _{m}$.

\subsection{Interaction D}

For the interaction~$Q=\beta _{0}\rho _{m}$ we introduce the new variable $z=%
\frac{\beta _{0}H_{0}}{H}$, such that the field equations read%
\begin{eqnarray}
\frac{dx}{d\tau } &=&\frac{1}{2}\left( \sqrt{6}\lambda y^{2}+3\left(
1+x^{2}+y^{2}\right) \right) +z\left( \frac{1-y^{2}}{x}+x\right) ,
\label{d.14} \\
\frac{dy}{d\tau } &=&\frac{1}{2}y\left( \sqrt{6}\lambda x+3\left(
1-x^{2}-y^{2}\right) \right) ,  \label{d.15} \\
\frac{dz}{d\tau } &=&\frac{3}{2}z\left( 1-x^{2}-y^{2}\right)  \label{d.16} \\
\frac{d\lambda }{d\tau } &=&\sqrt{6}\lambda ^{2}x\left( \Gamma \left(
\lambda \right) -1\right) ,~\Gamma \left( \lambda \right) =\frac{V_{,\phi
\phi }V}{V_{,\phi }^{2}}\text{\thinspace }.  \label{d.17}
\end{eqnarray}%
We remark that for this interaction, in order to write the dynamic system in
autonomous form, we should introduce a new variable such that to increase
the dimension. Similarly, before, we consider the exponential potential $%
V\left( \phi \right) =V_{0}e^{\lambda \phi }$, such that $\lambda $ is
always a constant.

The stationary points $D=\left( x\left( D\right) ,y\left( D\right) ,z\left(
D\right) \right) $ for the three-dimensional dynamical system (\ref{d.14}), (%
\ref{d.15}) and (\ref{d.16}) are%
\begin{eqnarray*}
D_{1}^{\pm } &=&\left( \pm 1,0,\frac{3}{2}\right) , \\
D_{2} &=&\left( \frac{\lambda }{\sqrt{6}},\sqrt{1+\frac{\lambda ^{2}}{6}}%
\right) , \\
D_{3}^{\pm } &=&\left( \pm i,0\right) \text{\thinspace }, \\
D_{4}^{\pm } &=&\frac{1}{\lambda }\sqrt{\frac{3}{2}}\left( -1,i\right) .
\end{eqnarray*}%
For the real points $D_{1}^{\pm }$ and $D_{2}$ we calculate the physical
quantities $\Omega _{m}$ and $q$, that is,%
\begin{eqnarray*}
D_{1}^{\pm } &:&\Omega _{m}=2,~q=-1, \\
D_{2} &:&\Omega _{m}=0,~q=-1-\frac{\lambda ^{2}}{2}.
\end{eqnarray*}

The eigenvalues of the linearized system for the real points are calculated 
\begin{eqnarray*}
D_{1}^{\pm } &:&-3,~-3,~\pm \sqrt{\frac{3}{2}}\lambda , \\
D_{2} &:&-\frac{1}{2}\left( 6+\lambda ^{2}\right) ,~-\left( 3+\lambda
^{2}\right) ,~-\frac{\lambda ^{2}}{2}.
\end{eqnarray*}

Stationary points $D_{1}^{\pm }$ describe de Sitter universes, where the
dark matter and the phantom field interact, point $D_{1}^{+}$ is an
attractor for $\lambda <0$; on the other hand, point $D_{1}^{-}$ is an
attractor for $\lambda >0$. Finally, point $D_{2}$ describes a universe
dominated by the phantom field; the physical properties are similar to that
of point $A_{3}$. Point $D_{2}$ is always an attractor. It is important to
mention that none of the stationary points depend on the parameter $\beta
_{0}$ value.

\begin{figure}[tbph]
\centering\includegraphics[width=0.5\textwidth]{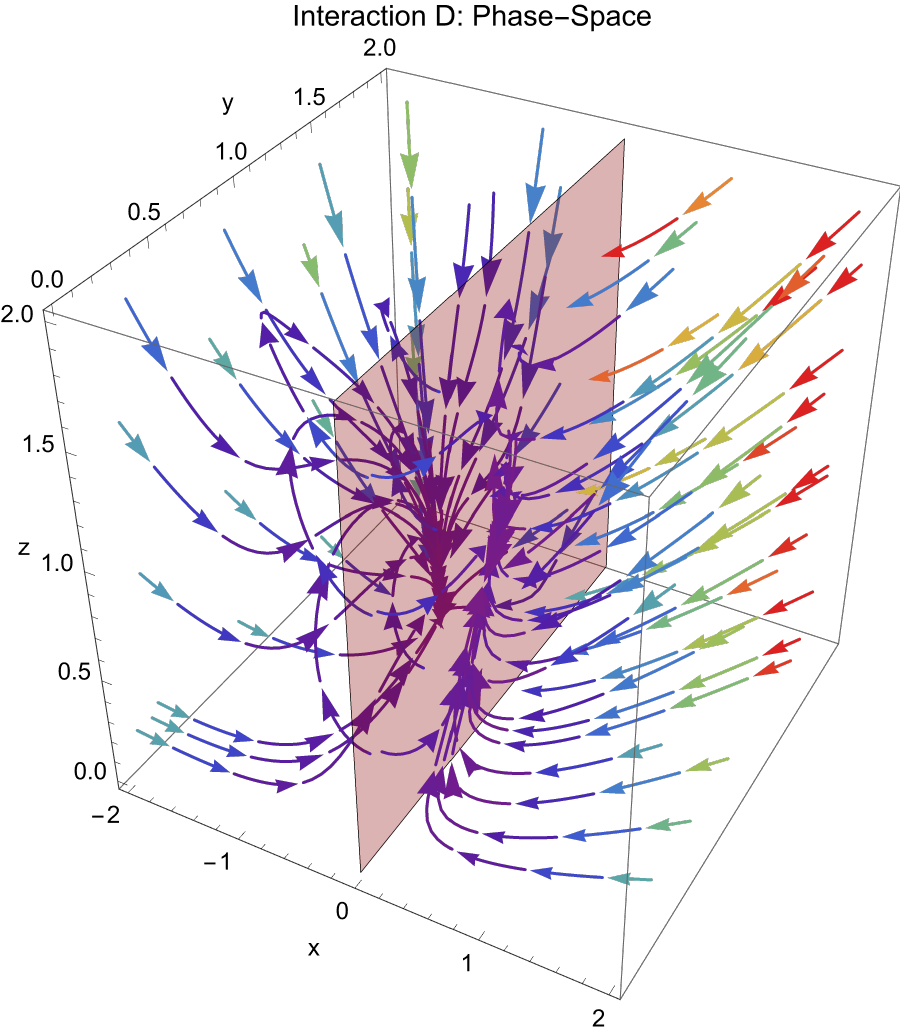}
\caption{Interaction D: Phase-space portrait for the dynamical system (%
\protect\ref{d.14}), (\protect\ref{d.15}) and (\protect\ref{d.17}). With
red, we marked the surface of singular points with $x=0$. }
\label{ff11}
\end{figure}

Similar to the model $C$, there is a singular surface $x=0$. The family of
points $D_{1}^{T}=\left( 0,y,0\right) $ and $D_{2}^{T}=\left( 0,1,z\right) $
are transition points for the interaction $C$, as discussed. In Figs \ref%
{ff10} and \ref{ff11} we present phase-space portraits for the dynamical
system (\ref{d.14}), (\ref{d.15}) and (\ref{d.16}). We observe that all the
trajectories reach the stationary points at the finite regime.

\begin{figure}[tbph]
\centering\includegraphics[width=1\textwidth]{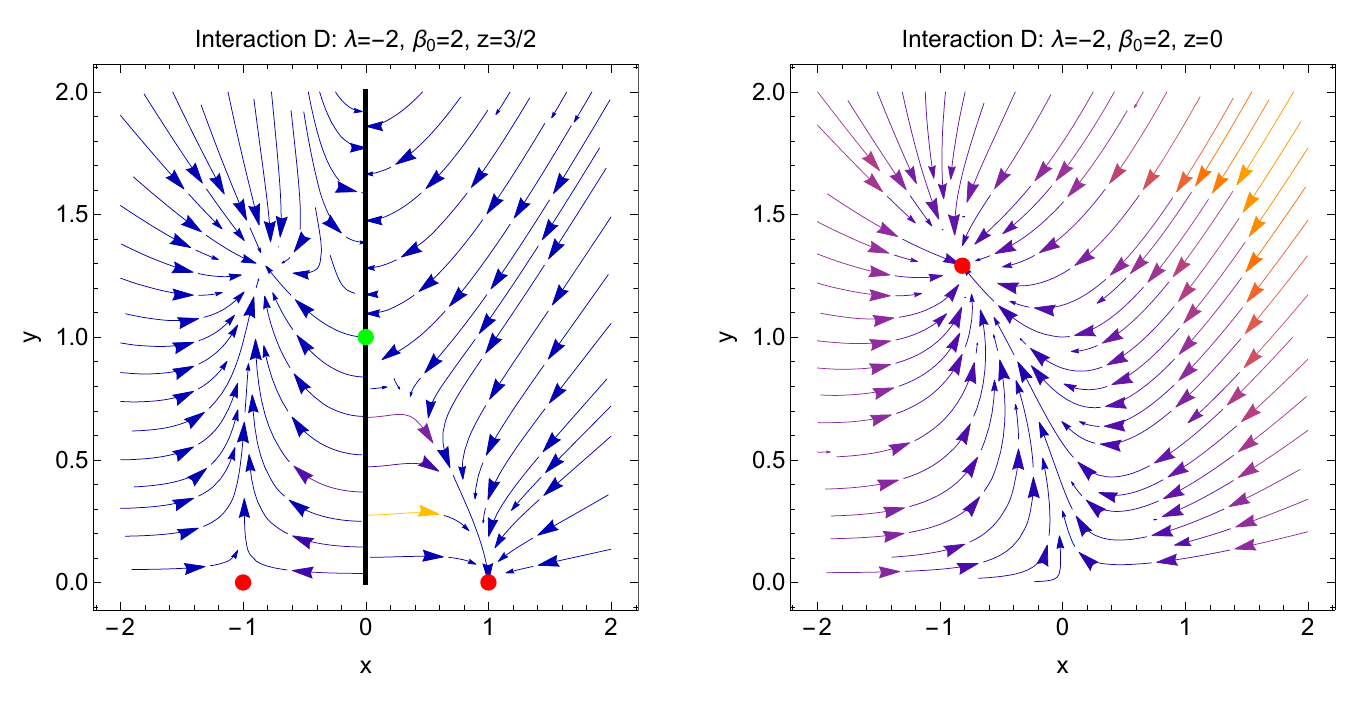}
\caption{Interaction D: Phase-space portraits for the dynamical system (%
\protect\ref{d.14}), (\protect\ref{d.15}) on the two surfaces $z=\frac{3}{2}$
Moreover, $z=0$ where the stationary points lie. With red are marked the
stationary points, while with green is marked the transition point $%
D_{T}^{2} $. The black line defines the singular points with $x=0$. }
\label{ff10}
\end{figure}

\subsubsection{Compactified variables}

\begin{figure}[tbph]
\centering\includegraphics[width=1\textwidth]{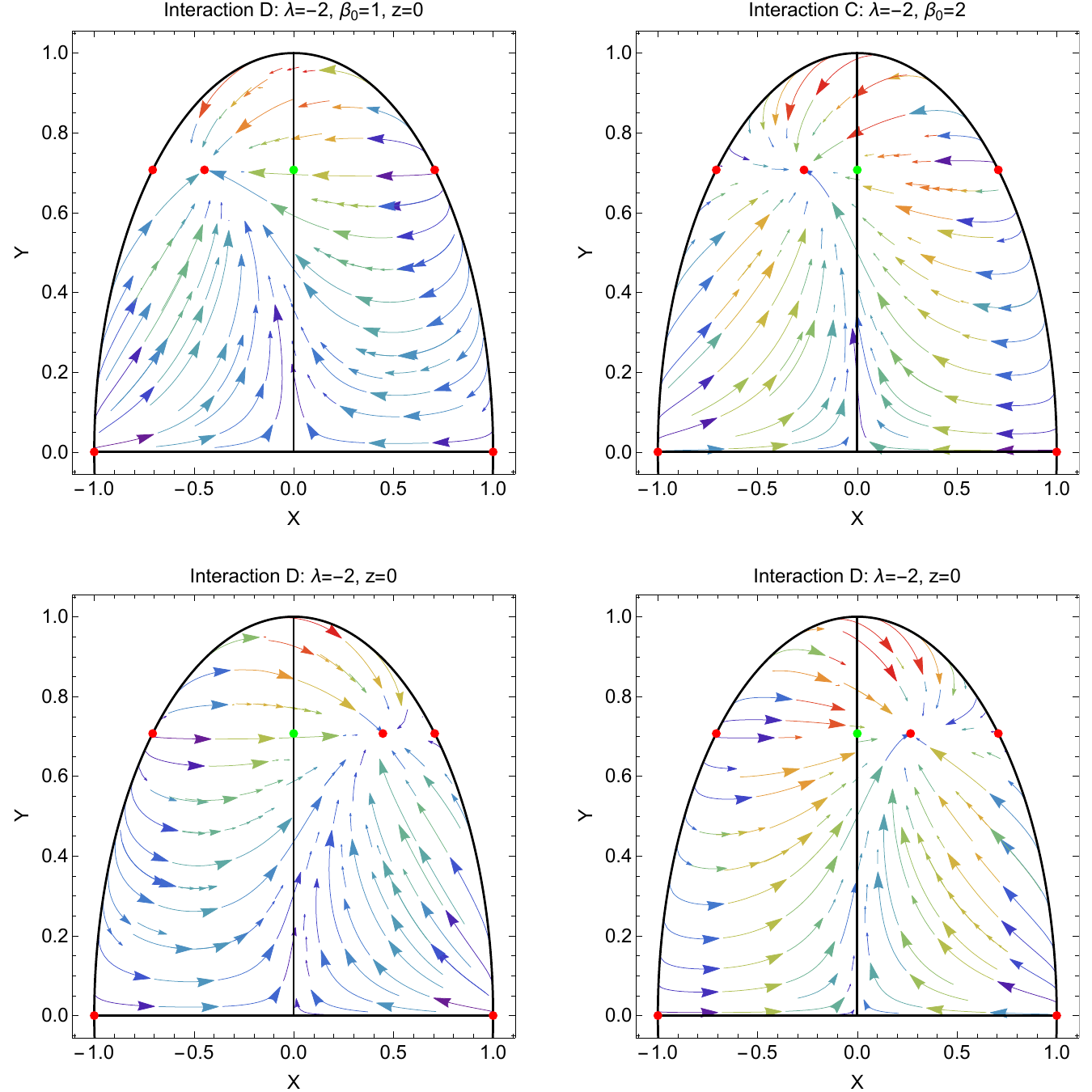}
\caption{Interaction D: Phase-space portraits for the dynamical system (%
\protect\ref{d.20}), (\protect\ref{d.21}), (\protect\ref{d.22}) for
different values of the free parameters $\left\{ \protect\beta _{0},\protect%
\lambda \right\} ~$on the surface $z=0$. The stationary points are marked
with red. }
\label{ff12}
\end{figure}

\begin{figure}[tbph]
\centering\includegraphics[width=1\textwidth]{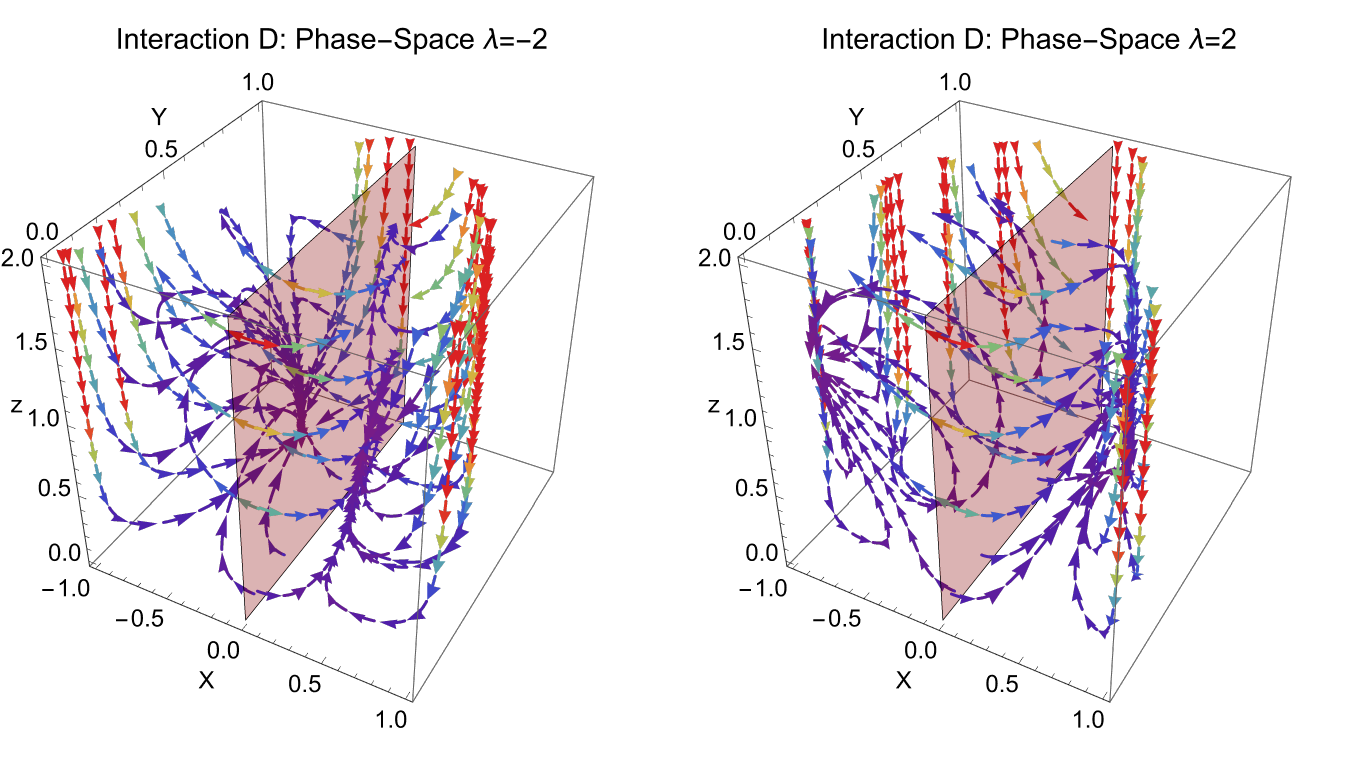}
\caption{Interaction D: Phase-space portraits for the dynamical system (%
\protect\ref{d.20}), (\protect\ref{d.21}), (\protect\ref{d.22}) for
different values of the free parameters $\left\{ \protect\beta _{0},\protect%
\lambda \right\} $.}
\label{ff14}
\end{figure}

In the compactified variables, the dynamical system reads%
\begin{align}
\frac{dX}{dT} &=\frac{1}{2}\left( \sqrt{6}\lambda Y^{2}\left(
1-2Y^{2}\right) -\left( 3X\left( 1+2Y^{2}\right) -\frac{\left(
1-X^{2}\right) \left( 1-2Y^{2}\right) }{X}z\right) \sqrt{1-X^{2}-Y^{2}}%
\right) ,  \label{d.20} \\
\frac{dY}{dT} &=\frac{1}{2}Y\left( 1-2Y^{2}\right) \left( \sqrt{6}\lambda
X+\left( 3-2z\right) \sqrt{1-X^{2}-Y^{2}}\right) ,  \label{d.21} \\
\frac{dz}{dT} &=\frac{3}{2}z\frac{1-2\left( X^{2}+Y^{2}\right) }{\sqrt{%
1-X^{2}-Y^{2}}}  \label{d.22}
\end{align}%
The latter dynamical system at infinity, i.e., on the surface $%
1-X^{2}-Y^{2}=0$, admits the following stationary points 
\begin{eqnarray*}
D_{1}^{\left( \infty \right) \pm } &=&\left( \pm 1,0,0\right) , \\
D_{2}^{\left( \infty \right) \pm } &=&\left( \pm \frac{1}{\sqrt{2}},\frac{1}{%
\sqrt{2}},0\right) .
\end{eqnarray*}%
The stationary points correspond to unstable solutions, which describe Big
Rip singularities. In Fig. \ref{ff12}, we present phase-space portraits on
the surface $z=0$, where the unique attractor is point $D_{2}$.
Three-dimensional phase-space diagrams are presented in Fig. \ref{ff14}.

\section{Beyond the exponential potential}

\label{sec4}

In the previous section, we investigated the stationary points in the case
of the exponential potential, which corresponds to a constant parameter $%
\lambda $. Nevertheless, for other functional forms for the potential
parameter $\lambda $ is varying, and a new family of stationary points
exists. \ In the following, we assume that the scalar field potential $%
V\left( \phi \right) $ leads to a smooth function $\Gamma \left( \lambda
\right) $. We present qualitative data on the admitted stationary points,
and we investigate the existence of new possible solutions. Since $\Gamma
\left( \lambda \right) $ is arbitrary, we do not study the stability
properties.

\subsection{Interaction A}

Consider the three-dimensional system (\ref{g.25}), (\ref{g.26}) and (\ref%
{g.27}). If $\bar{A}=\left( x\left( \bar{A}\right) ,y\left( \bar{A}\right)
,\lambda \left( \bar{A}\right) \right) $ is a stationary point, then%
\begin{eqnarray*}
0 &=&\sqrt{6}ax^{2}-3x^{3}-3x^{2}\left( 1+y^{2}\right) +\sqrt{6}\left( \beta
_{0}+\left( \lambda -\beta _{0}\right) y^{2}\right) , \\
0 &=&y\left( \sqrt{6}\lambda x+3\left( 1-x^{2}-y^{2}\right) \right) , \\
0 &=&x\lambda ^{2}\left( \Gamma \left( \lambda \right) -1\right) .
\end{eqnarray*}%
If $\lambda _{0}$ is a solution for the algebraic equation $\lambda
_{0}^{2}\left( \Gamma \left( \lambda _{0}\right) -1\right) =0$, then we
recover all the stationary points for the exponential potential. However,
from the last equation, there exists a new family of stationary points with $%
x=0.$ Hence, it follows that $\lambda =0$ and $y=1$, which is nothing else
than point $\bar{A}=\left( A_{3}\left( \lambda \rightarrow 0\right)
,0\right) $. Hence, there is not any new family of solutions provided by
other functions for the potential.

\subsection{Interaction B}

We apply the same procedure for the dynamical system (\ref{d.03}), (\ref%
{d.04}) and (\ref{d.05}) in order to determine the stationary points $\bar{B}%
=\left( x\left( B\right) ,y\left( B\right) ,\lambda \left( B\right) \right) $%
. For $\lambda _{0}$ which solves the equation $\lambda _{0}^{2}\left(
\Gamma \left( \lambda _{0}\right) -1\right) =0$, we recover the stationary
points given by the exponential potential. However, we determine a new
family of stationary points with coordinates $\bar{B}=\left( 0,0,\lambda
\right) $, where $\lambda $ is arbitrary. These points generalize the
matter-dominated solution described by point $B_{1}.$ Thus, other forms of
the potential provide no new families of physical solutions.

\subsection{Interaction C}

The dynamical system (\ref{d.08}), (\ref{d.09}) and (\ref{d.10}) is well
defined for $x\neq 0$, thus no new families of solutions exist for other
forms of the scalar field potential.

\subsection{Interaction D}

The dynamical system for the interacting model D is well defined for $x\neq
0.$ Thus, the same conclusion as that of the interacting model C follows.

\section{Conclusions}

\label{sec5}

In this study, we investigated the cosmological dynamics of a cosmological
model with the interaction between dark energy and dark matter. We
considered the dark energy to be described by a phantom scalar field. The
phantom field violates the weak energy conditions, and it can have a
negative energy density, while the equation of state parameter can cross the
phantom divide line.

We considered the simplest interaction function, $Q=\beta \left( t\right)
\rho _{m}$, where we introduced four different forms for the coefficient
function $\beta \left( t\right) $, specifically we studied the models with
(A) $\beta \left( t\right) =\beta _{0}\dot{\phi}$,~(B) $\beta \left(
t\right) =\beta _{0}\frac{\dot{\phi}^{2}}{H}$, (C) $\beta \left( t\right)
=\beta _{0}H$ and (D)~$\beta \left( t\right) =\beta _{0}H_{0}$. The
definition for the function $\beta \left( t\right) $ leads to different
interacting models, consequently to different cosmological histories. In
order to examine the global phase space, we make use of compactified
variables and we study for the existence of stationary points at the
infinity regime. Independent of the interacting model, there exist two sets
of stationery points that describe Big Rip singularities; however, the
stability properties of these points depend on the nature of the
interaction. We summarize the results which found for the case of the
exponential scalar field potential.

Model A leads to a set of field equations which admit three stationary
points at finite regime, two of the points, namely $A_{1}$ and $A_{2}$
describe the coexistence between the two components of the cosmological
fluid, while the third point describes a universe dominated by the phantom
field. Possible attractors are points $A_{1}$ and $A_{3}$ while the Big Rip
singularities are avoided due to the appearance of the interaction; this is
in agreement with the study presented in \cite{cop2}.

Model B admits four stationary points at the finite regime. Point $B_{1}$
describes a matter-dominated era, point $B_{2}$ a scalar field dominated
era, while $B_{3}$ and $B_{4}$ describe universes where the two elements for
the cosmological fluid contributes, and there exists a nonzero interaction
component. Contrary to before, the Big Rip singularities can be attractors.
Nevertheless, they can avoided when the coupling constant $\beta _{0}<\frac{1%
}{2}$, where in this case the unique attractor is point $B_{2}$, and points $%
B_{3}$,~$B_{4}$ does not exist.

Models C and D introduce a singularity in the field equations when the
scalar field is constant, i.e. $\dot{\phi}=0$. This model admits another set
of characteristic points: transition points where the dynamical parameter $%
\dot{\phi}$ can change sign in the phase-space. The interacting model, $C$,
admits three stationary points at the finite regime, two of the points $C_{1}
$ and $C_{2}$ describe a nonzero interacting component, and point~$C_{3}$
corresponds to a universe dominated by the phantom field. The possible
attractors for this model are point $C_{1}$ and $C_{3}$. Finally, the
interacting model, $D$, possesses two stationary points at the finite
regime. Points $D_{1}^{\pm }$ describe de Sitter solutions with a nonzero
interaction, and point $D_{2}$ describe a phantom field-dominated universe.
The unique attractor is the scaling solution given by point $D_{2}$. Due to
the existence of the singular surfaces for the interacting models C and D,
someone should be very careful when imposing initial conditions to determine
numerically the evolution of the physical parameters.

From this analysis, we remark that interaction A, provided by theories with
variational principles, leads to a universe where Big Rip singularities can
always be avoided; however, there is no epoch where only the dark matter
source contributes to the universe. On the other hand, interacting model~B
provides the matter-dominated era, with the downside of the Big Rip
singularities to be possible future attractors. Nevertheless, from the
phase-space analysis, we observe that it is possible to constrain the
initial condition problem to avoid future cosmic singularities. Finally,\
the interacting models C and D suffer from the singular surfaces on the
phase space, which makes it not preferred for describing the cosmological
history.

Last but not least, we discussed the effects of an arbitrary scalar field
potential function on the physical properties of the asymptotic solutions at
the stationary points.

\begin{acknowledgments}
AP \& GL thanks the support of Vicerrector\~{A}-a de Investigaci\~{A}%
${{}^3}$%
n y Desarrollo Tecnol\~{A}%
${{}^3}$%
gico (VRIDT) at Universidad Cat\~{A}%
${{}^3}$%
lica del Norte (UCN) through Resoluci\'{o}n VRIDT No. 096/2022 and Resoluci%
\'{o}n VRIDT No. 098/2022. This study was funded by FONDECYT 1240514, Etapa
2024.
\end{acknowledgments}

\end{document}